\documentclass[twocolumn]{aastex63}

\newcommand{\sindex}{$S$}

\received{}
\revised{}
\accepted{}
\submitjournal{ApJL}

\shorttitle{Active-region nesting on solar-like stars}
\shortauthors{I\c{s}{\i}k et al.}
\begin{document}

\title{Amplification of brightness variability by active-region nesting in solar-like stars}

%% The \author command is the same as before except it now takes an optional
%% argument which is the 16 digit ORCID. The syntax is:
%% \author[xxxx-xxxx-xxxx-xxxx]{Author Name}
%%
%% The new \altaffiliation can be used to indicate some secondary information
%% such as fellowships. This command produces a non-numeric footnote that is
%% set away from the numeric \affiliation footnotes.  NOTE that if an
%% \altaffiliation command is used it must come BEFORE the \affiliation call,
%% right after the \author command, in order to place the footnotes in
%% the proper location.
%%
%% Use \email to set provide email addresses. Each \email will appear on its
%% own line so you can put multiple email address in one \email call. A new
%% \correspondingauthor command is available in V6.3 to identify the
%% corresponding author of the manuscript. It is the author's responsibility
%% to make sure this name is also in the author list.
%%
%% While authors can be grouped inside the same \author and \affiliation
%% commands it is better to have a single author for each. This allows for
%% one to exploit all the new benefits and should make book-keeping easier.
%%
%% If done correctly the peer review system will be able to
%% automatically put the author and affiliation information from the manuscript
%% and save the corresponding author the trouble of entering it by hand.

%\correspondingauthor{Emre I\c{s}{\i}k}

\author[0000-0001-6163-0653]{Emre I\c{s}{\i}k}
\altaffiliation{Current address: Dept. of Computer Science, Turkish-German University,
Istanbul, Turkey}
\affiliation{Max-Planck-Institut f\"ur Sonnensystemforschung, Justus-von-Liebig-Weg 3, 37077 G\"ottingen, Germany}
\affiliation{Dept. of Computer Science, Turkish-German University, 
\c{S}ahinkaya Cd. 108, Beykoz, 34820 Istanbul, Turkey}
\affiliation{Feza G\"ursey Center for Physics and Mathematics, Bo\u{g}azi\c{c}i University, Kuleli 34684 Istanbul, Turkey}
\email{emre.isik@tau.edu.tr}

\author[0000-0002-8842-5403]{Alexander I. Shapiro}
\affiliation{Max-Planck-Institut f\"ur Sonnensystemforschung, Justus-von-Liebig-Weg 3, 37077 G\"ottingen, Germany}

\author[0000-0002-3418-8449]{Sami K. Solanki}
\affiliation{Max-Planck-Institut f\"ur Sonnensystemforschung, Justus-von-Liebig-Weg 3, 37077 G\"ottingen, Germany}
\email{solanki@mps.mpg.de}
%\altaffiliation{Max-Planck-Institut f\"ur Sonnensystemforschung  % \\
%Justus-von-Liebig-Weg 3 % \\
%37077, G\"ottingen, Germany}
\affiliation{School of Space Research, Kyung Hee University, Yongin, Gyeonggi-Do, 
	      446-701, Republic of Korea}

\author[0000-0002-1377-3067]{Natalie A. Krivova}
\affiliation{Max-Planck-Institut f\"ur Sonnensystemforschung, Justus-von-Liebig-Weg 3, 37077 G\"ottingen, Germany}

\begin{abstract}

\emph{Kepler} observations revealed that hundreds of stars with near-solar fundamental parameters and rotation periods have much stronger and more regular brightness variations than the Sun.
Here we identify one possible reason for the peculiar behaviour of these stars.  Inspired by solar nests of activity, we assume that the degree of inhomogeneity of active-region (AR) emergence on such stars is higher than on the Sun. To test our hypothesis, we model stellar light curves by injecting ARs consisting of spots and faculae on stellar surfaces at various rates and nesting patterns, using solar AR properties and differential rotation. We show that a moderate increase of the emergence frequency from the solar value combined with the increase of the degree of nesting can explain the full range of observed amplitudes of variability of Sun-like stars with nearly the solar rotation period. 
Furthermore, nesting in the form of active longitudes, in which ARs tend to emerge in the vicinity of two longitudes separated by $180^\circ$, leads to highly regular, almost sine-like variability patterns, rather similar to those observed in a number of solar-like stars.
\end{abstract}

%\keywords{stars: activity --- 
%stars: solar-type --- stars: rotation --- (stars:) starspots}

\section{Introduction} \label{sec:intro}

Rotational brightness variability of Sun-like stars results from the transit of stellar magnetic features, i.e. dark spots and bright faculae, over the visible stellar hemisphere.
The advent of high-precision photometry brought by planet-hunting missions allowed measuring rotational variability for several hundred thousands of stars \citep{timo13,wb13, basri13, mcquillan14}. It also took solar-stellar connection studies to a new level, by allowing solar variability to be compared with that of solar peers. %-like stars, i.e. stars with near-solar 
%fundamental parameters and {\it known} near-solar rotation rates.

Recently, \cite{timo20} (hereafter R20) identified 2529 stars with near-solar fundamental parameters, but with unknown rotation periods (hereafter, pseudosolar stars) and 369 stars with near-solar rotation periods and near-solar fundamental parameters (hereafter, solar-like stars). While many of the pseudosolar stars are expected to have near-solar rotation periods, the sample most probably also includes stars with different periods (in particular, stars rotating slower than the Sun). 
The light curves of pseudosolar stars appear rather irregular, often resembling the solar light curve, and their variability amplitudes lie
mostly within the solar range. Consequently, the Sun appears to be a typical star belonging to the pseudosolar sample: it would most probably be attributed to this sample if observed by {\it Kepler}, owing to difficulties in determining its rotation period).

Surprisingly, the variability pattern of solar-like stars differs from that of the pseudosolar stars and the Sun. Firstly, R20 found that the mean variability in their solar-like sample was 5 times stronger than %the median 
solar variability. Secondly, many of the stellar light curves had a rather regular temporal profile, often resembling a sine wave. This is in stark contrast to the Sun, which has a complex and quite irregular light curve. Especially at periods of high magnetic activity the solar light curve becomes so irregular that even the solar rotation period cannot be determined correctly from photometry, using standard methods like the auto-correlation function and Lomb-Scargle periodograms \citep[see, e.g.][]{Aigrain2015,Veronika2020, Eliana2020}. 

We illustrate the difference between the variability of the Sun and solar-like stars in Fig. 1, where we compare the {\it Kepler} light curve of KIC 7019978 ($T_{\rm eff}=5726$~K, $P_{\rm rot}=26.76$~d) with that of the Sun during the maximum of Cycle 23.
The solar light curve as it would be observed in the {\it Kepler} passband has been taken from \cite{nina20}. 
The strong dip in the solar light curve (t$\sim$ 2950~d) was caused by a rather unusual configuration of magnetic features on the solar surface: three very large active regions (hereafter ARs) on one longitudinal hemisphere (i.e., within a longitudinal extent of $180^\circ$) had emerged contemporaneously, while the 
opposite hemisphere remained free of ARs. Single dips 
of such depths occur rarely the solar brightness variations. 
The remaining shallower dips represent the typical effect of sunspot groups around activity maximum. 
The rotational brightness modulation of \object{KIC~7019978} is clearly more 
vigorous and much more regular than that of the Sun, which varies its brightness on a comparable level only during the transit of that extraordinary concentration of three very large ARs. 

 %Furthermore, \citet{Zhang20} found that these stars, including the Sun, have systematically lower activity levels than the periodic sample of R20.
All in all, the R20 study raised an intriguing question: how can large-amplitude and regular light curves of solar-like stars be produced and, in particular, what is the difference between the surface magnetism of the Sun and of solar-like stars? 
We introduce here two potential hypotheses for explaining the difference between the Sun and solar-like stars: 
$(i)$ the surface coverage of magnetic features is larger than 
typical solar levels, leading to higher variability; 
$(ii)$ the longitudinal distribution of magnetic features is more clumpy than on the Sun, 
leading to an amplification of brightness variations for a given level of magnetic flux or activity. 
%(iii)$ active regions and in particular starspots live longer than on the Sun, and/or have a flatter size distribution than on the Sun, also leading to a more periodic signal. 
%In this study, we consider the first two of these possibilities in forward-modelling of 
%brightness variability. 

The first hypothesis is supported by the fact that solar cycles can have
different strengths. For example, cycle 19 was stronger than Cycle 24 by a factor of 
three to four in terms of the maximum average sunspot number. One can speculate that the dynamos of solar-like stars can produce cycles at a wider range of strengths than those of the Sun, although it is \emph{a priori} unclear why this should be so. 

The second hypothesis is motivated by the tendency of solar ARs to emerge into 
regions of recent magnetic flux emergence, called 
complexes or nests of activity \citep{gaizauskas83,castenmiller86,brouwer90,usoskin05}.
The exact physical mechanisms for AR nesting are as yet unclear. 
The weak rotational variability and the irregular light variations of the Sun (and pseudosolar stars with undetermined rotation periods) could be related to the low degree of active-region nesting \citep[e.g., only 40-50\% of sunspots 
are associated with nests, see][]{pc02}. On solar-like stars with high variability and regular light curves, magnetic flux emergence 
can be clustered to higher degrees than on the Sun. 
The effects of AR nesting on photometric variability has not been 
studied to date. 
Here we present a simple
model that quantifies the effects of nesting on photometric variability of solar-like stars in the \emph{Kepler} 
passband for a range of magnetic activity levels. 
%Possible physical mechanisms underlying 
%active-region nesting are beyond the scope of the present study, so we simply assume 
%that nesting can occur over a range of degrees and in different modes. 

\begin{figure}
\centering
\includegraphics[width=\columnwidth]{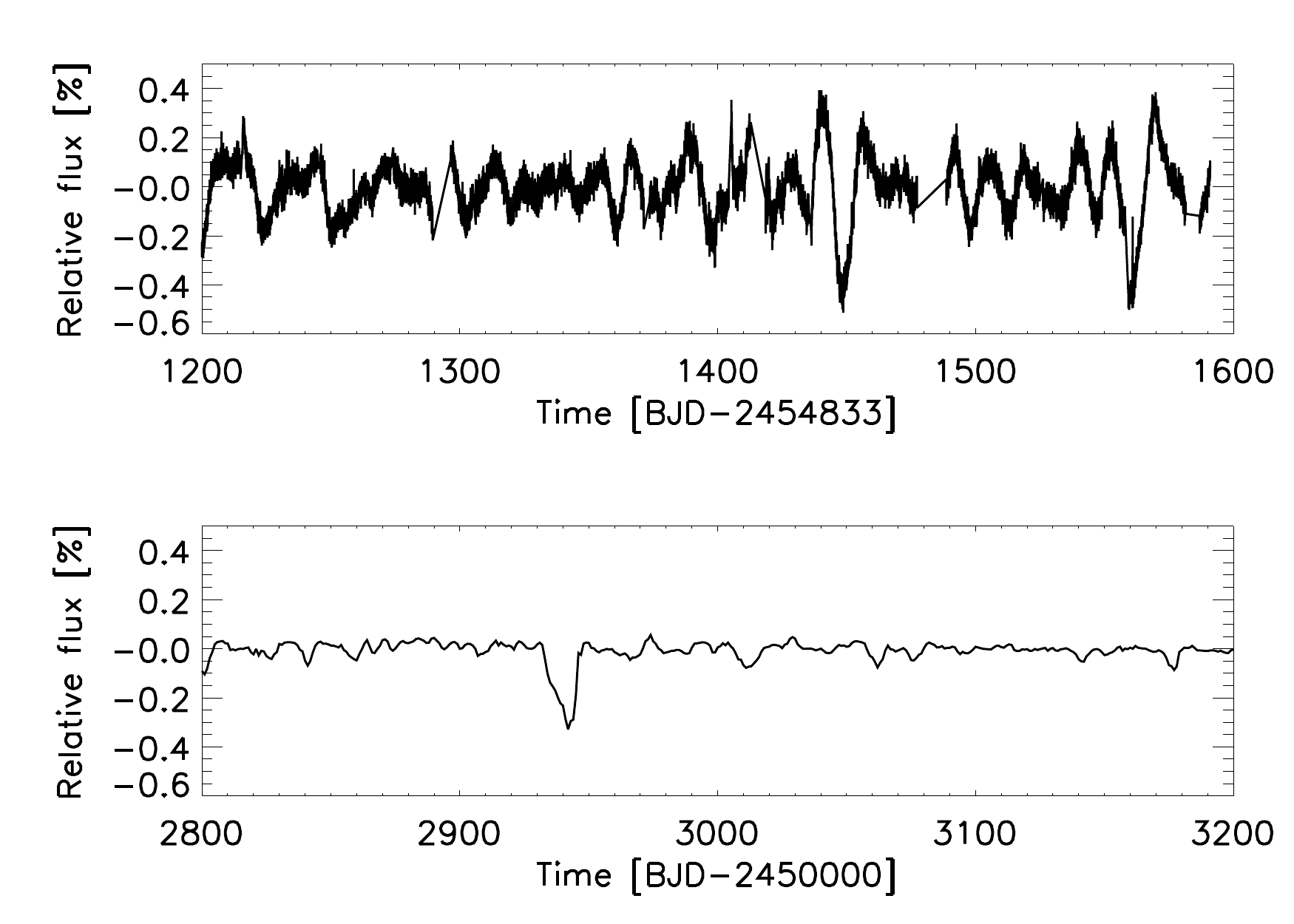}
\includegraphics[width=.5\columnwidth]{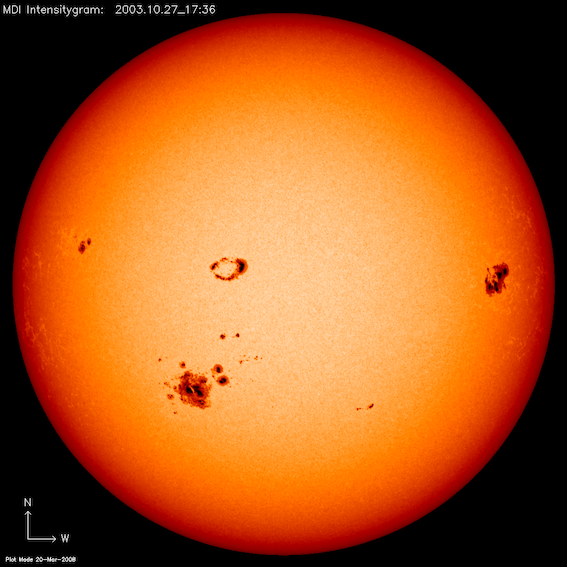}
\caption{Brightness variations of KIC 7019978 (top panel) and of 
the Sun (middle panel), plotted on the same scale of the vertical axis. The SOHO/MDI image (bottom panel) corresponds to the time of the 
exceptionally strong dip in the solar light curve above, around 2940 days. 
Note the conspicuous differences in shape and amplitude of the variations 
of the two stars. }
\label{fig:examples}
\end{figure}

\begin{figure*}
\centering
\includegraphics[width=.32\linewidth]{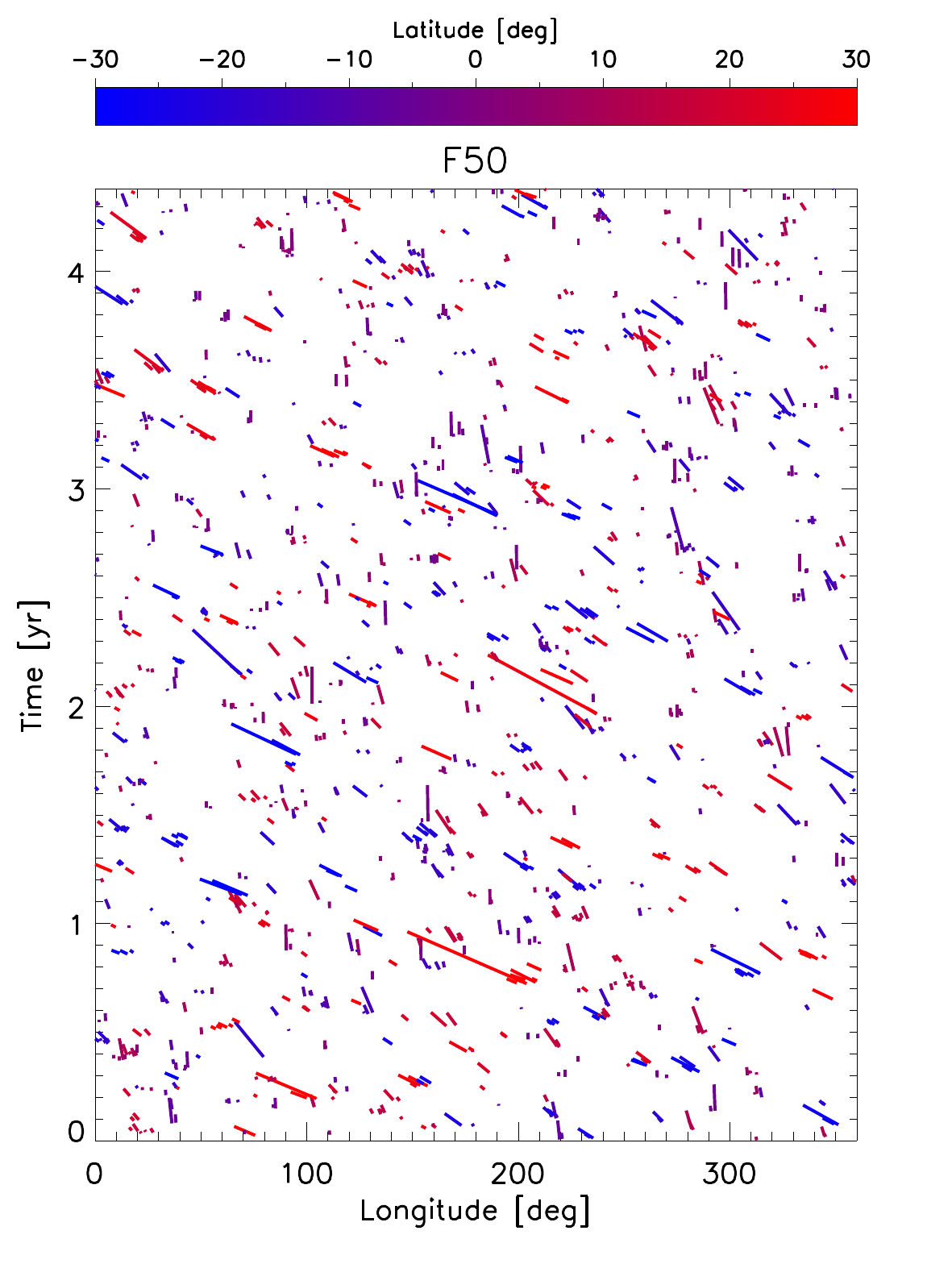}
\includegraphics[width=.32\linewidth]{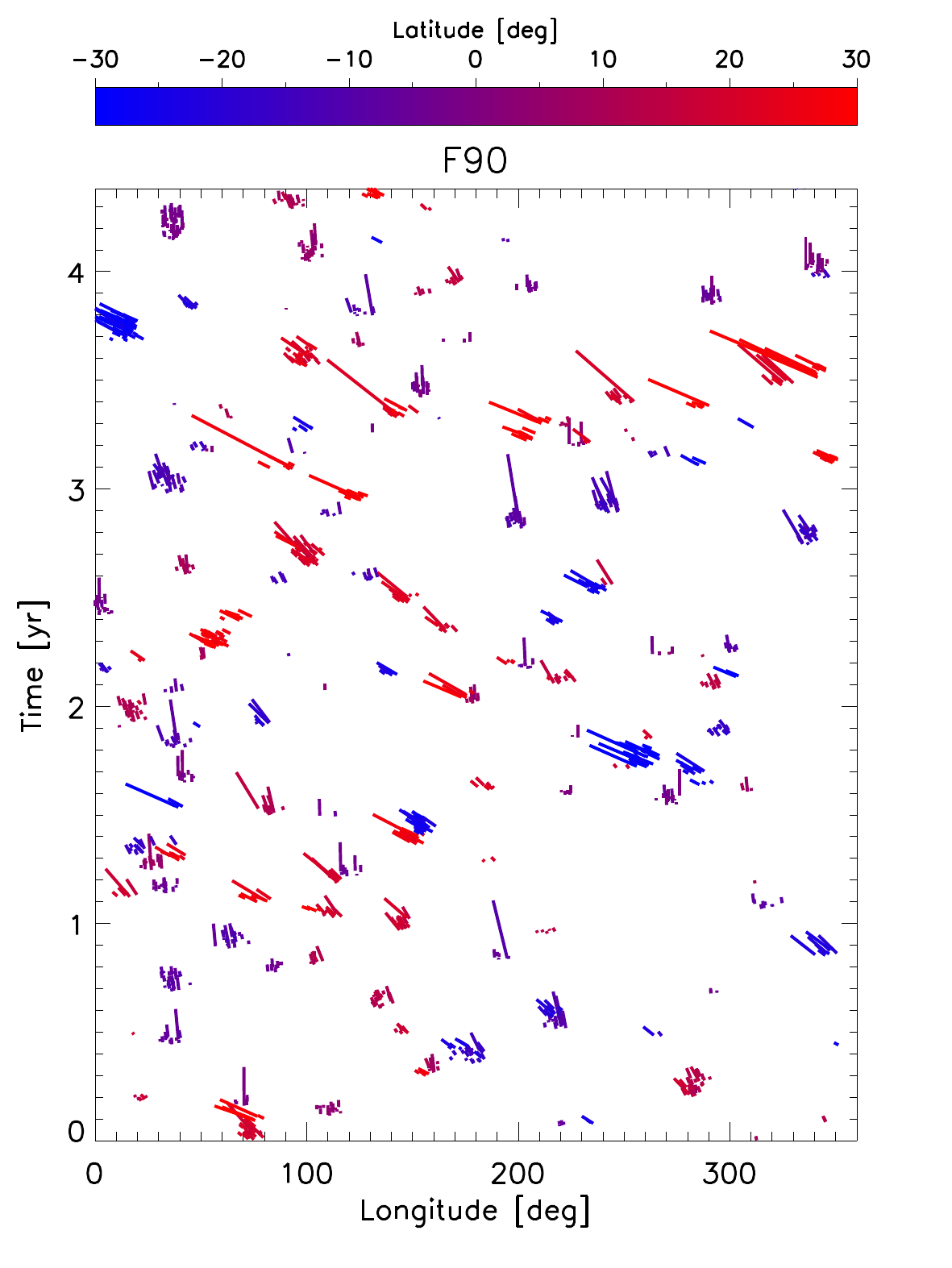}
\includegraphics[width=.32\linewidth]{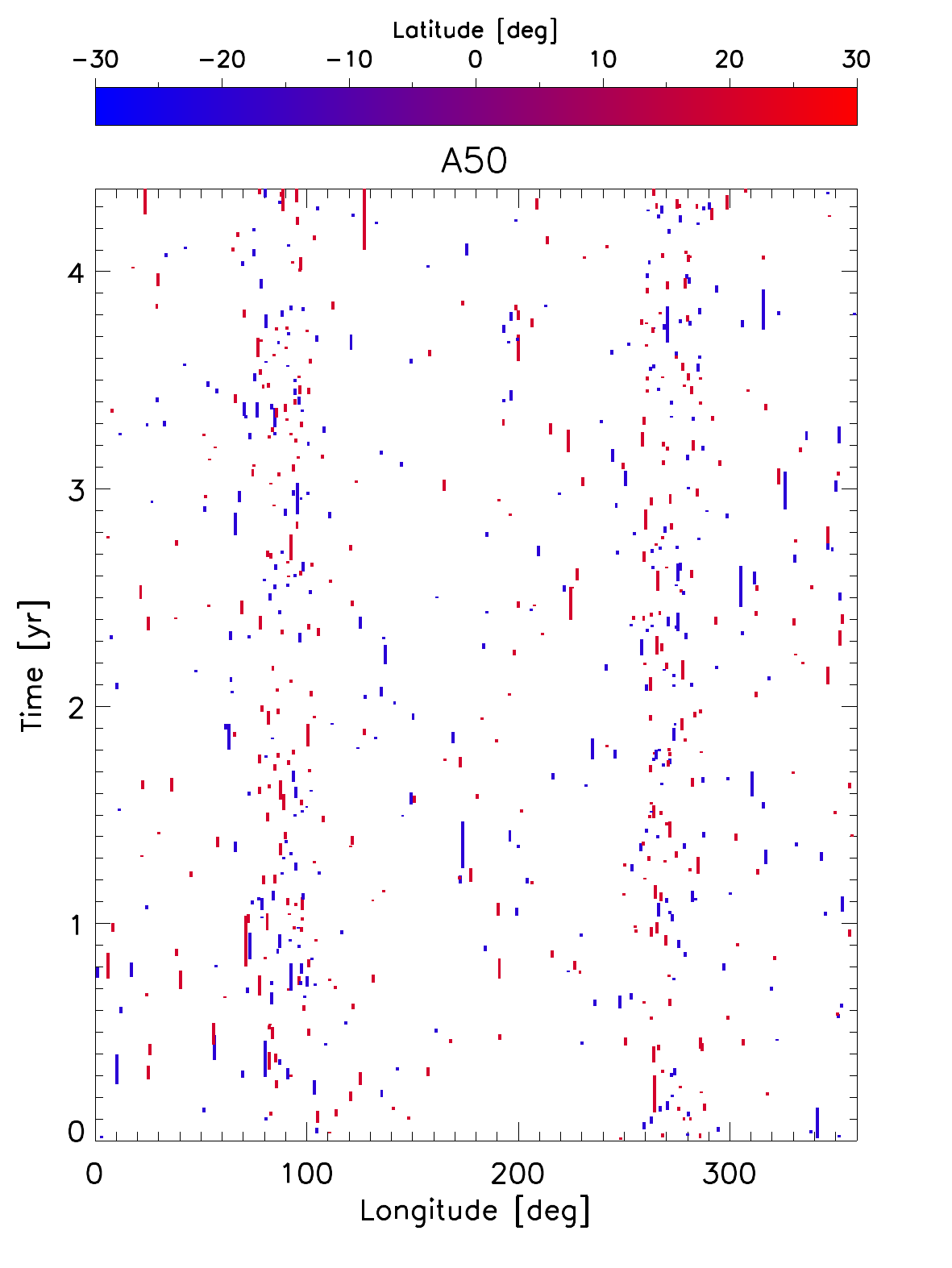}
\caption{Time-longitude maps of active-region occurrence for an activity level close to that of a typical solar maximum, \sindex$=0.182$. The panels show the nesting modes F50, F90, and A50, from left to right (see main text for definitions). Colours denote AR latitudes. 
The vertical extension of each element represents the spot-group lifetime for the corresponding 
AR. }
\label{fig:nestest}
\end{figure*}

\newpage
\section{Method}
\label{sec:method}

\subsection{Modelling brightness variations}
\label{ssec:ingred}

To simulate stellar light curves, we employ the model developed by 
 \citet{shapiro20}.
We simulate brightness variations of a solar-like star as observed equator-on (inclination $i=90^\circ$) for a period of 
4.4 years (close to \emph{Kepler}'s operation time) 
with a sampling rate of four measurements per day. 
Different activity levels are obtained by letting different numbers of ARs (consisting of spot groups and faculae) emerge within the latitude limits $\pm 30^\circ$, assuming a size distribution of ARs as on the Sun \citep[i.e., a log-normal function;][]{bogdan88,bs05}. 
The AR emergence frequency depends on the activity level, but for a given activity level it is constant in time, so that we do not consider any systematic change in the activity level during the simulated period. Following the approach of \cite{shapiro20}, we use solar dependences of facular and spot disc coverages on the chromospheric
$S$-index \citep[given by Eqs.~1--2 from][]{shapiro14}. We thus quantify the activity level in terms of $S$, which is taken as an average over the simulated time range. We consider a linear (in time) decay law of active regions and fixed the ratio between the lifetime of facular and spot parts of ARs to three.
A more detailed description of light-curve modelling is given in Appendix~\ref{sec:a1}. 

To measure variability, we split the entire light curve into 90-day segments (to be compatible with \emph{Kepler} quarters), calculate the peak-to-peak variability in each of them, and take the mean value among the segments, which we call $R_{90}$. 
We consider peak-to-peak variability instead of 95\% to 5\% difference often employed in the literature \citep[e.g.,][R20]{basri13}, because our simulated light curves are free of noise. 

\begin{figure*}
\centering
\includegraphics[width=\columnwidth]{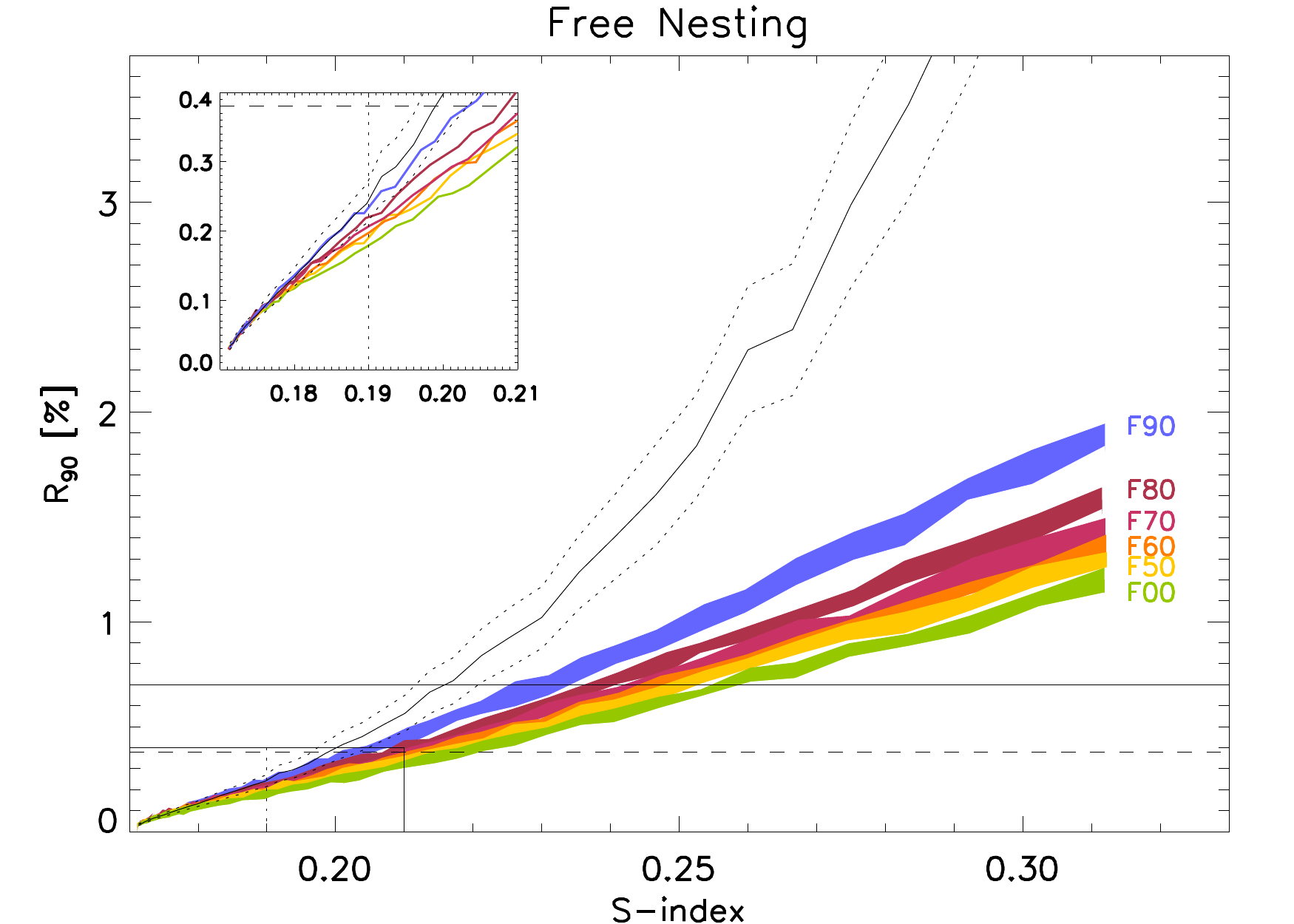}
\includegraphics[width=\columnwidth]{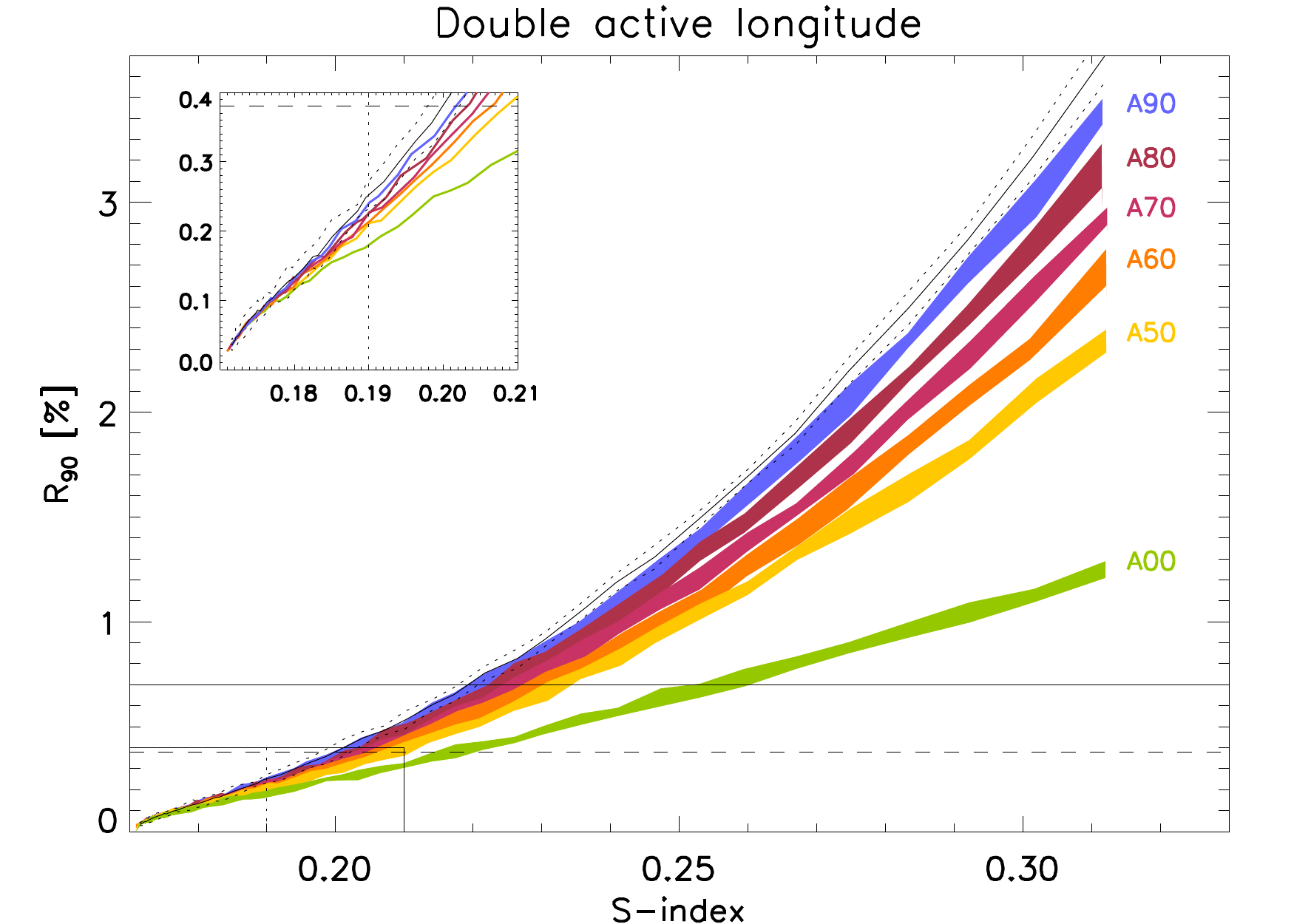}
\caption{Mean amplitudes of rotational variability as a function of 
\sindex~for free nesting (left) and active longitudes (right). The vertical thickness of each coloured band represents the $\pm 1\sigma$ scatter 
in $R_{90}$ for a set of 10 random realisations (for F100, 50 realisations were used). The standard deviation in \sindex~is of order $10^{-4}$. 
The black curves show the case of 100\% nesting (F100 and A100), with $1\sigma$ levels shown by dashed curves. The inset depicts the part of the main plot indicated in the 
lower left corner (solid rectangle). The vertical dashed line at \sindex$=0.19$ represents the maximum of the annually averaged solar S-index observed until now. 
The full and long-dashed horizontal lines mark the maximum and the mean variability in the solar-like sample of R20, respectively. 
{ Note that the calculations are performed for stars observed from their equatorial planes; thus they represent upper limits of the rotational variability. }
}
\label{fig:LCvar-25}
\end{figure*}

\subsection{Nesting active regions}
\label{ssec:nesting}

To simulate nests of ARs, we consider two distinct modes. 
In the \emph{free-nesting} mode, clusters of magnetic features form and decay sequentially. 
An active region (AR) emerges either (a) in the vicinity of 
the previous emergence location with a fixed probability $p$, or (b)
at a random location within the latitude range of $\pm 30^\circ$ with the remaining probability $1-p$. 
The procedure is very similar to the one introduced by \citet[][see their Appendix C]{isks18}. 
The proximity of an AR's emergence location to the nest centre is drawn from a 
two-dimensional Gaussian process with standard deviations of 
$2^\circ$ in latitude and $3^\circ$ in longitude, which closely represents the 
observational values given by \citet{pc02}. The algorithm is designed in such a way 
that multiple nests can form sequentially, 
not contemporaneously. This means that before all the members of a given nest have fully 
emerged, a new nest does not begin to form. The development of a nest is thus completed 
when the subsequently emerging AR does not belong to the nest.
However, a new nest can start to form when 
an existing nest (or nests) have not fully decayed, owing to size-dependent lifetimes of 
constituent spots and faculae.

In the \emph{double-active-longitude} (AL) mode, the probabilistic procedure is the same as for the free-nest mode, but ARs tend to emerge near one of the two `active' longitudes separated by $180^\circ$ (with equal probability). The proximity of an AR to its host active-longitude is set randomly, using a normal distribution with $\sigma=10^\circ$ in longitude. 

In both nesting modes we set the probability $p$ of a given AR to be associated to a nest 
(whether a free-nest or AL) as a free parameter. 
In the following, we will use the notations F$P$ and A$P$, where the prefix describes 
the nesting mode (F for free nesting and A for AL), and $P$ stands for the 
percentile probability of a given AR to emerge as part of a nest. 

In the free nesting case, the nest centres, as well as ARs emerged inside or outside of nests follow the solar 
surface differential rotation over latitude $\lambda$ \citep{snodgrass83}:
\begin{equation}
\Omega^\prime(\lambda) = - 2.3\sin^2\lambda - 1.62\sin^4\lambda~~~{\rm degrees/day},
\label{eq:difrot}
\end{equation}
where $\Omega^\prime = \Omega(\lambda)-\Omega(0)$ is the rate at which 
spots drift in longitude (in the equatorial rest-frame).

On the Sun, ALs are detected only in a dynamical reference frame, synchronised with the latitude-dependent rotation periods of major flux emergence regions \citep{usoskin05}. Even if we assume an initially coherent AL, differential rotation would destroy this pattern in only a few stellar rotations. Taking a rigid AL and applying differential rotation to individual ARs on smaller scales would have only a minor effect on the variability. To keep our model simple, we thus opted for rigid rotation in the AL mode, to compute brightness variations for an extremely coherent mode of nesting, because our goal is to assess the entire potential variability range. 

Figure~\ref{fig:nestest} shows the locations and lifetimes of spot groups for models 
F50, F90, and A50, for \sindex$=0.182$, typical 
of a solar maximum. 
Longer lifetimes are associated with larger spot groups, in agreement with the 
Gnevyshev-Waldmeier rule of sunspot groups. 
From F50 to F90 there is a dramatic increase in the 
degree of clustering of spot groups, which is expected to amplify photometric variability. Also the A50 case demonstrates significant clustering in longitude.
In the F50 and F90 cases the differential rotation leads spot groups and free nests to exhibit longitudinal drifts at rates 
depending on the latitude \citep[see, e.g.,][]{ozavci18}. 

\section{Results}
\label{sec:res}

\subsection{Nesting amplifies variability}
\label{ssec:freepar}

\begin{figure*}
\centering
\frame{\includegraphics[width=\columnwidth]{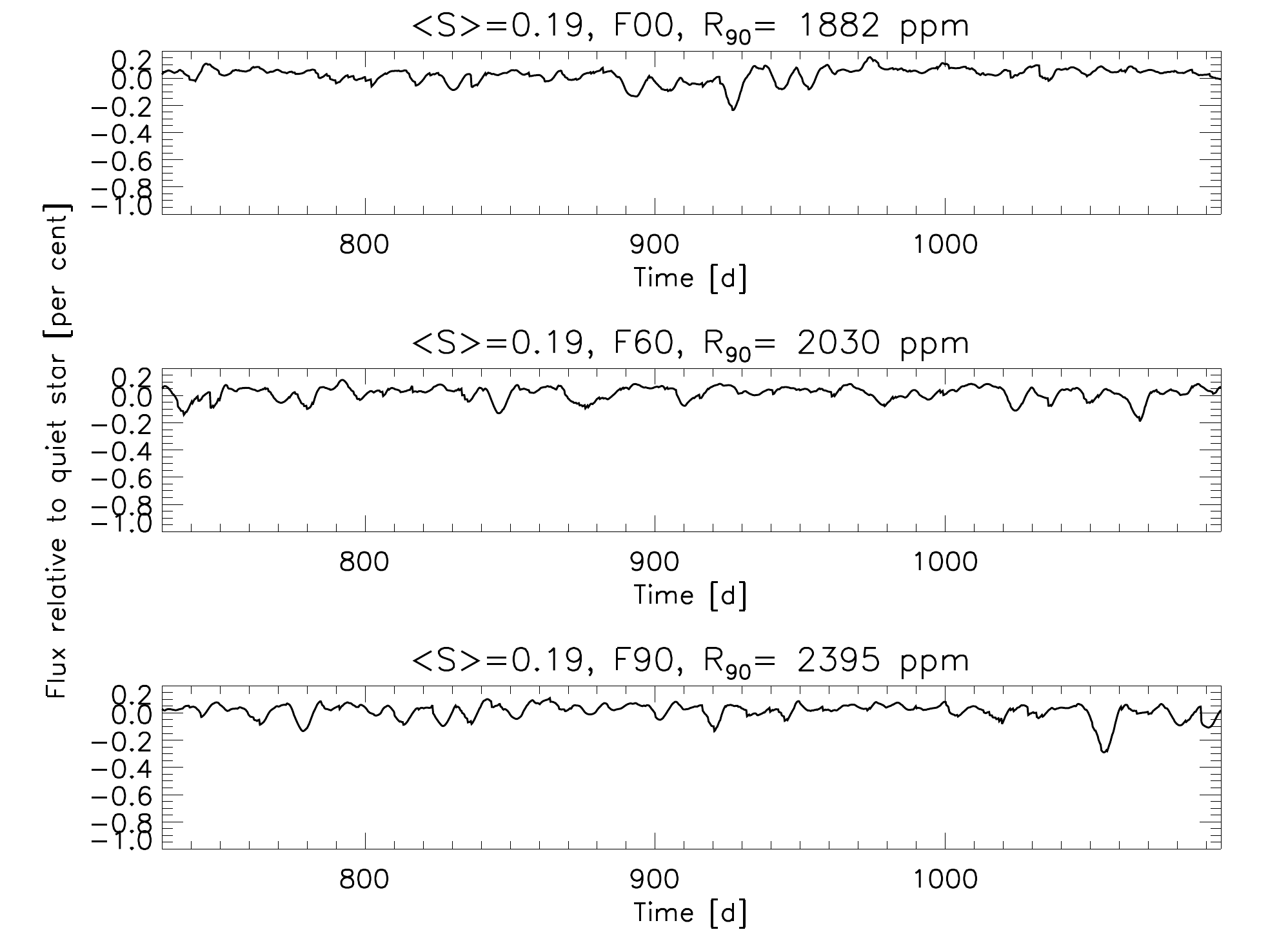}
\includegraphics[width=\columnwidth]{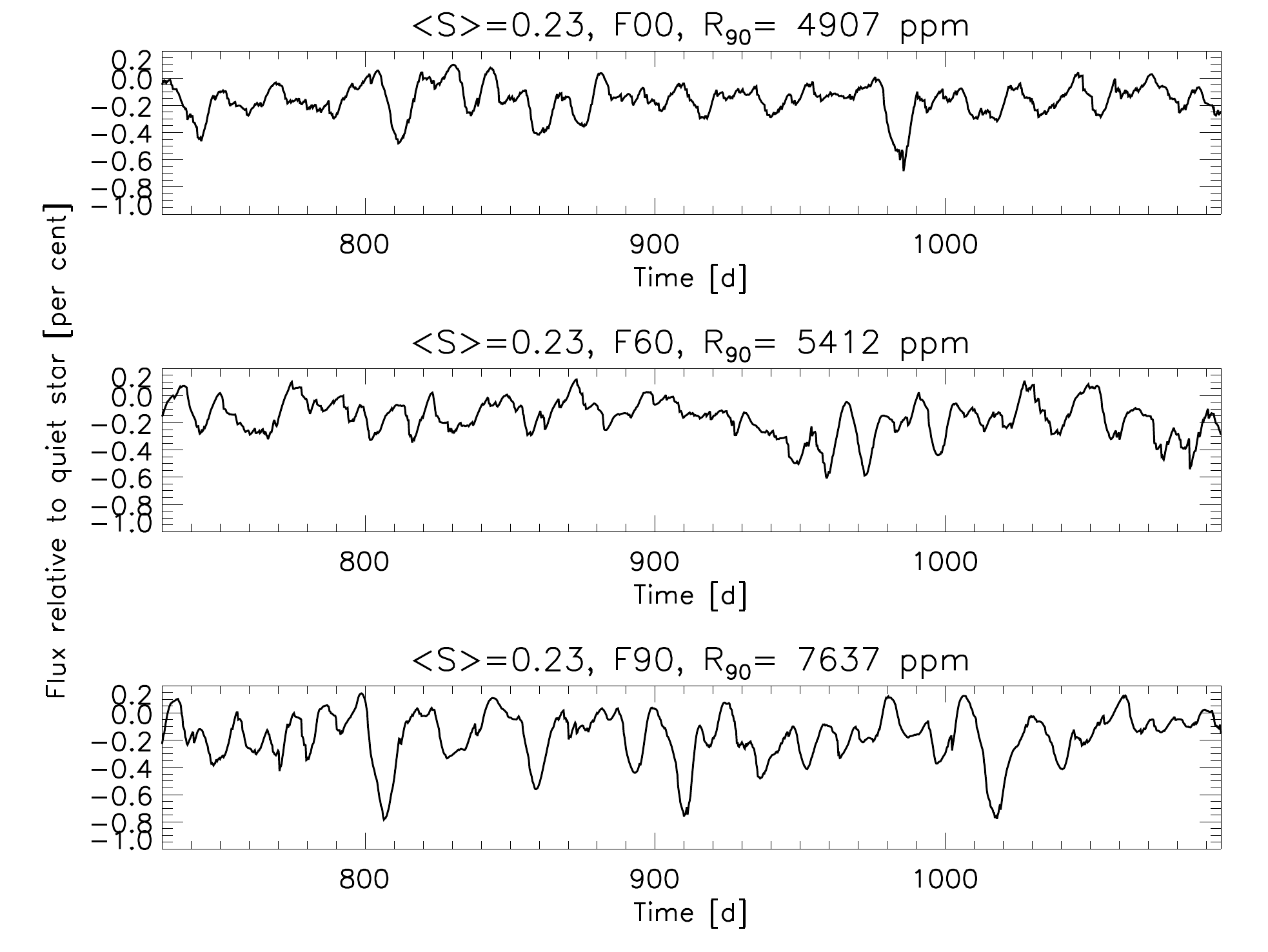}} \\
\frame{\includegraphics[width=\columnwidth]{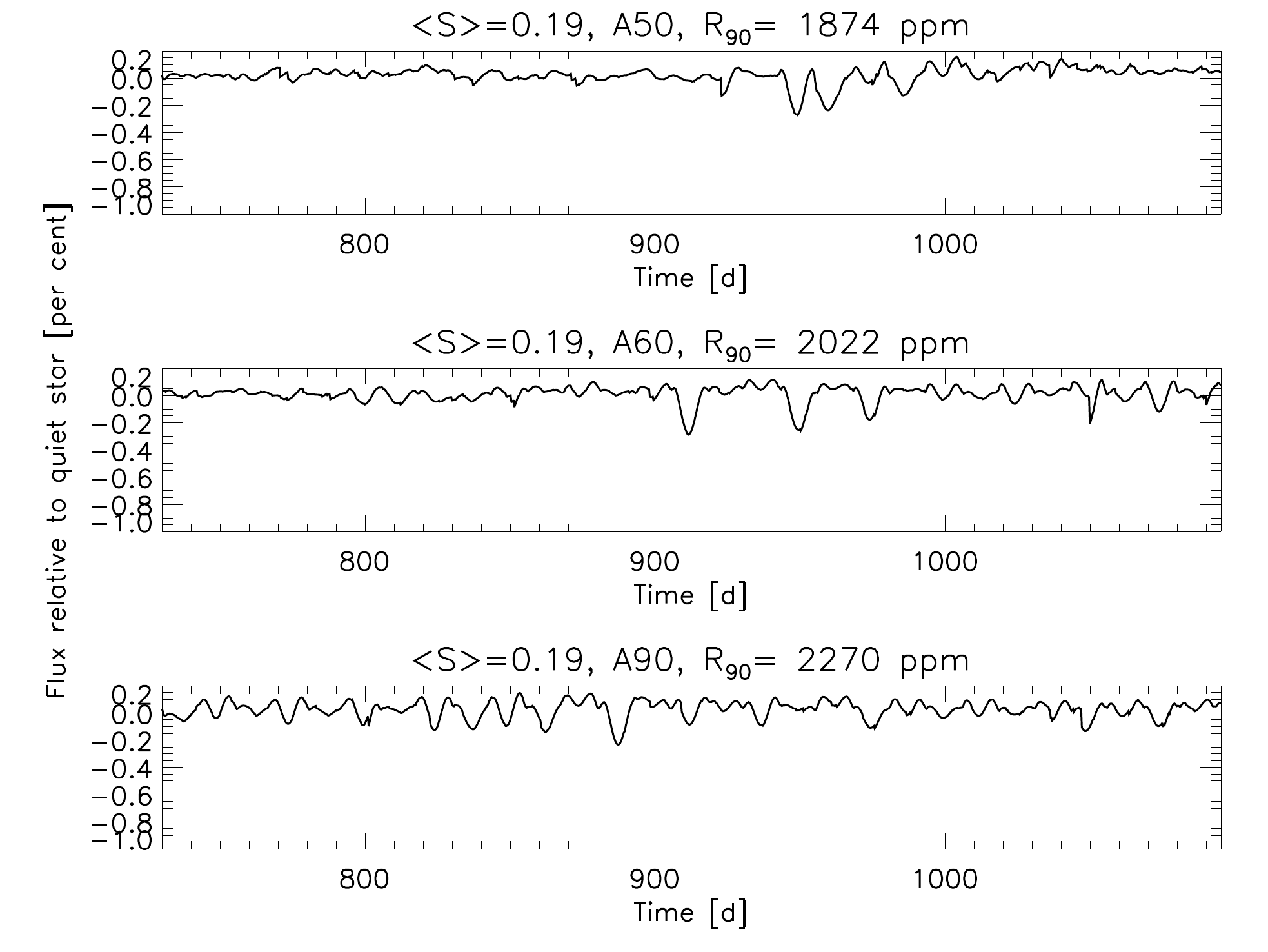}
\includegraphics[width=\columnwidth]{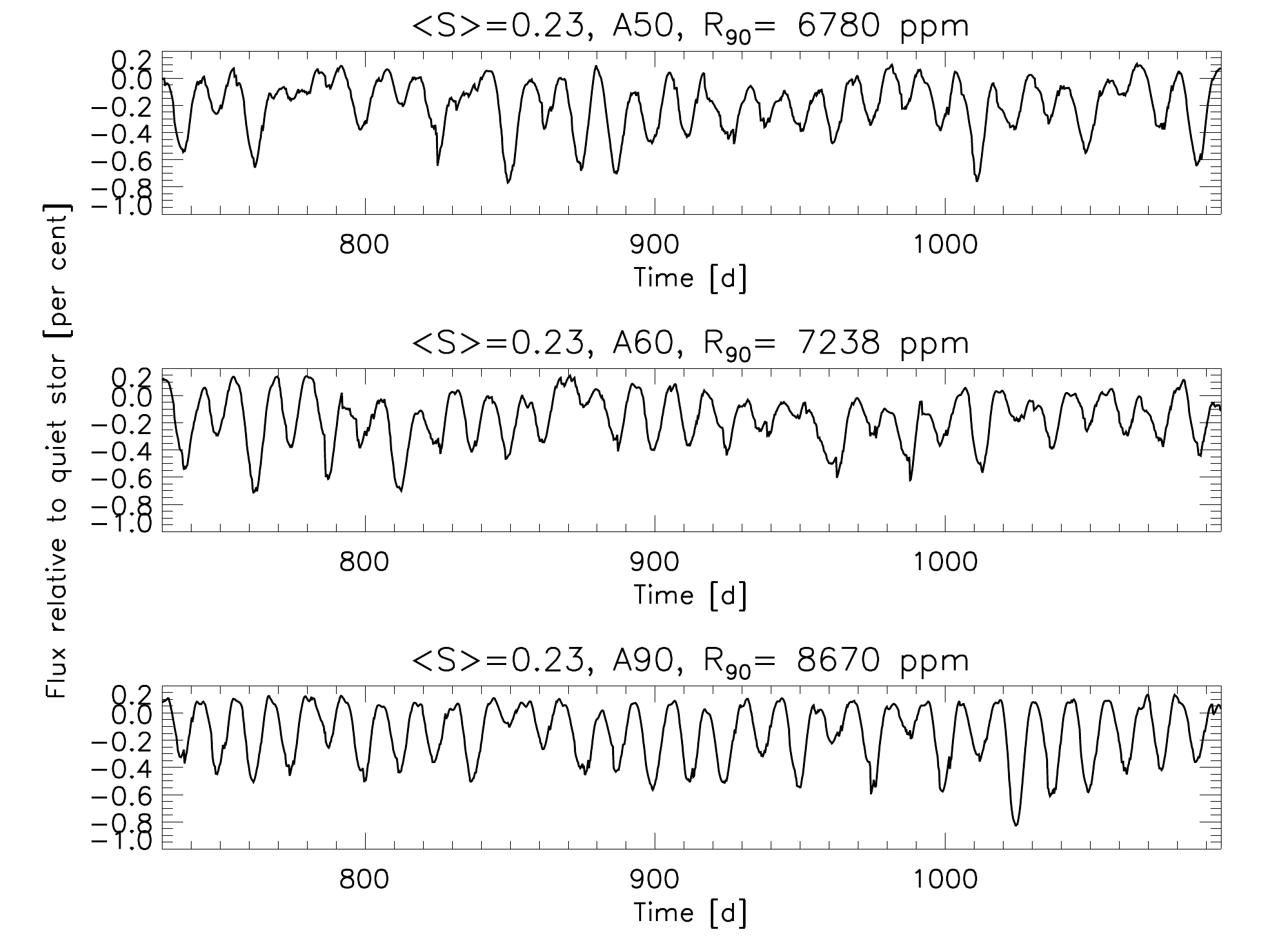}}
%\frame{\includegraphics[width=.5\columnwidth]{Lcurve-free-s19-dif}
%\includegraphics[width=.5\columnwidth]{Lcurve-free-s23-dif}} \\
%\frame{\includegraphics[width=.5\columnwidth]{Lcurve-fixem2-s19-rig}
%\includegraphics[width=.5\columnwidth]{Lcurve-fixem2-s23-rig}}
\caption{Sample light curves for free-nesting (upper panels) and active-longitude (lower 
panels) modes. For each mode, two activity levels (columns) are shown with three degrees of 
nesting each (rows). 
$S\simeq 0.19$ in the left panels and $S\simeq 0.23$ on the right panels. 
The nesting mode and degree, and the 25-day variation 
amplitude, $R_{90}$, are indicated above each panel. Note that for $S=0.19$ the slightly higher $R_{90}$ of the case F90 compared to A90 is a mere coincidence and does not contradict the overall trend of the AL mode yielding higher variability than free-nest mode (see Fig.~\ref{fig:LCvar-25}).}
\label{fig:LCfree}
\end{figure*}

We first investigate the dependence of the photometric variability on the level of magnetic activity
for different modes and degrees of nesting. 
{ Since we are interested in finding the upper limit of variability, we only performed simulations for stars observed from their equatorial planes. Solar-like stars observed from out of this plane will show smaller variabilities \citep{nina20}. }
We carried out 10 random realisations of AR emergences for each combination of 
activity level and degree of nesting, with the exception of 50 realisations for F100. { Such multiple runs were 
made to quantify the posterior width owing to randomness 
in AR emergence. }

Figure~\ref{fig:LCvar-25} shows $R_{90}$ 
as a function of \sindex~for free-nesting and active-longitude modes. 
Variability increases with \sindex~in all cases, owing to an increasing occurrence of ARs (which depends on the \sindex, see Sect.~\ref{sec:method}). For F00 (i.e., for random emergence locations), $R_{90}$ scales roughly linearly with \sindex. The dependences for higher degrees of nesting gradually deviate from linearity with increasing \sindex. 
This is because the area coverage of spots is proportional to \sindex$^2$ in the model, in accordance with solar data. 
A random distribution of spots would thus imply that the variability 
is proportional to the square-root of spot coverage, i.e., to \sindex. 
Stronger nesting steepens the dependence and in the extreme case of 100\% nesting the variability is proportional to the coverage, i.e. to \sindex$^2$. 

Comparison of the left and right panels of Fig.~\ref{fig:LCvar-25} shows that the AL mode enhances variability even more than free nesting. This is
because the permanent active longitudes lead to a more coherent superposition of ARs during each full rotation. %contributions to variability. 
Starting from A50, the dependences are nearly quadratic, 
in parallel with the functional relationship between the spot coverage and \sindex.
  %The curve for 
%A100 (100\% nesting) is comparable with P100, though it yields systematically lower variabilities, because 
%the same number of ARs now emerge into two ALs in anti-phase. 

In their sample of solar-like stars, 
R20 detected variability amplitudes of up to 0.7\%.
Taking $R_{90}=0.7\%$ as an upper limit, 
we can consider various possibilities for how such a value can be reached, using Fig.~\ref{fig:LCvar-25}. For the solar level of nesting (F50) an activity level of $S=0.25$ is needed. According to Eq.~(1-2) from \cite{shapiro14} employed in our model (see Sect.~2), such an S-index corresponds to mean spot and facular coverages of about 17 and 5 times larger than the corresponding annually averaged solar values during the maximum of Cycle 22.
For F100 (a single nest throughout four years), the variability
$R_{90}=0.7\%$ is reached already at $S\simeq 0.215$ (spot and facular coverages are about 6 and 2 times higher than solar, respectively). 
For F90 the same level is reached at $S\simeq 0.23$ (spot and facular coverages 10 and 3 times solar).

In the active-longitude mode, $R_{90}=0.7\%$ is reached for A50 at $S\simeq 0.23$, similar to F90. For A90, $S\simeq 0.22$ would be sufficient (spot and facular coverages 7 and 2.5 times solar, respectively). The significant difference between the extreme cases F100 and A100 is because in free nesting F100 forms a single nest, whereas A100 distributes the same number of emergences into two ALs in opposition, resulting in weaker variability.

\subsection{Active longitudes: more regular light curves}
\label{ssec:lc}

We now investigate the impact of nesting on the morphology of light curves. Figure~\ref{fig:LCfree} shows sample light curves, for $S\simeq 0.19$ 
(expected during a solar maximum) 
and $S\simeq 0.23$.
In the free-nest mode, the variability increases with the nesting degree (in line with Fig.~\ref{fig:LCvar-25}), but the brightness variation is still far from being sinusoidal, even for F90. In the active-longitude mode at $S=0.23$, the light curves not only reach the highest observed variability levels, but also become much more regular. They look very similar to the light curves shown by many periodic stars, i.e. sine-like, exhibiting smooth changes between consecutive dips.
Additional simulations involving a single AL with the same width resulted in stronger but less sinusoidal rotational modulations.

\section{Discussion and Conclusions}
Data collected by the {\it Kepler} space telescope indicate that stars with near-solar rotation rates and fundamental parameters can have a broad range of photometric variabilities. R20 have recently found that the periodic subsample of such stars appear to be significantly more variable than the Sun.

We have shown here that the full range of observed variabilities of such stars can be explained by preserving solar properties of active regions but going well beyond the degree of active-region nesting presently observed on the Sun and somewhat beyond the solar magnetic activity level.
This is accomplished using a very simple model, in which the fundamental properties of solar variability are reasonably well-represented, though for other stars some of the assumptions involved may not apply. 

{Our model allows connecting the difference between the rotational variability of pseudosolar and solar-like samples of R20 with the differences in the activity level. We have shown here that the exact connection depends on the degree of AR nesting. }
Our calculations indicate that the mean {and maximum} variability amplitudes of the solar-like sample analysed by R20 ($R_{90}$ { being} 0.38\% { and 0.70\%, respectively}) can be obtained by a simple increase of activity from the solar value (i.e. without increase of nesting). This, however, would require an increase of the activity level up to $S=0.22$ { for the mean and $S=0.25$ for the maximum}.
With AR nesting, say F90 case (compared to F50 for the Sun), we need a much more moderate increase of activity, $S\simeq 0.20$ { (mean) and $S\simeq 0.23$ (maximum)}). 

{ Both estimates appear to be} consistent with the recent
results from \citet[][{see their Fig.~1}]{Zhang20}, who utilised LAMOST
spectra to measure S-{ indices} of stars from the R20 sample and showed that { the} periodic, { `solar-like'} stars are moderately more active than the Sun.
{ Unfortunately, we cannot make a direct comparison of the S-indices found by 
\citet{Zhang20} with our numerical experiments, because the LAMOST S-index scale cannot be directly converted to the Mount-Wilson (MW) S-index used in our study. 
At the same time, Fig.~1 from \citet{Zhang20} shows that the mean S-index of the solar-like stars is lower than that of stars with rotation periods between 10 and 20 days. These faster-rotating stars are expected to have S-indices (in MW scale) between about 0.20-0.28 \citep[see Figs.~1-2 from ][]{shapiroa20}. }

Furthermore, we found that the nearly sinusoidal light curves of strongly variable solar-like stars can be reproduced by a highly non-axisymmetric pattern of AR emergence, with an azimuthal wavenumber of two. Consequently, such a pattern of AR emergence can simultaneously explain both the amplitude and morphology of the observed light curves. We note that double active longitudes have often been suggested to explain the light curves of very active BY Dra-, RS CVn- and FK Com-type cool stars \citep{berd05}. Most such stars are components of close binaries. 

We suggest three possible mechanisms for solar-like stars to exhibit active longitudes: 1) they can be components of non-eclipsing close binaries or they have warm or hot gas-giant components \citep{shibata13}. In 
the active components of such systems, tidal forces can lead to preferred longitudinal zones for emerging flux tubes \citep[see, e.g.,][though they found this tendency for 
$P_{\rm rot}\sim 2$~d]{voho03}. In that case, AL-type nesting can permanently lead to strong rotational variability \citep[see also][]{korhonen05}. 2) AL-type nesting can be a generic phenomenon, which all solar-type stars can undergo from time to time, or above some activity level. 3) Solar-like stars can have time-varying degree of nesting, as well as temporarily existing ALs, which can be responsible for those light curves exhibiting transitions between very regular and less regular variations. These possibilities should be explored more thoroughly in the future. 

The model presented in this letter was kept very simple, to demonstrate the overall effects of AR-nesting on brightness variability 
in solar-like stars. Physically more consistent models that simulate flux emergence and surface transport processes \citep[following][]{isks18} including effects of AR-nesting and the rotation rate are planned for a subsequent study. 

\acknowledgements
We thank T. Reinhold for providing the processed light curve of KIC 7019978 and useful discussions. { We also acknowledge the referee, whose requests improved the discussion of our results}. AS acknowledges funding from the European Research Council under the European Union 
Horizon 2020 research and innovation programme (grant agreement No. 715947). This work 
has been partially supported by the BK21 plus program through the National Research 
Foundation (NRF) funded by the Ministry of Education of Korea. The authors thank the International Space Science Institute, Bern, for their support of Science Team 446 and the resulting helpful discussions.

\appendix

\section{Forward modelling of light curves}
\label{sec:a1}
The detailed procedure of light-curve synthesis has been described by 
\citet{shapiro20}. We briefly explain in the following the main concept of the model relevant to our purpose of simulating rotational variability. 

In our model active regions (ARs) emerge instantaneously at random times within 1600 days. Consequently, the emergence frequency of ARs is constant, i.e. we do not consider stellar activity cycles. The activity level of a simulated star is then defined by the total number of emergences. For each realisation of emergences we calculate not only the time series of stellar brightness, but also the time series of disc area coverage by spots. Using the dependence between the stellar disc coverage by spots and the $S$-index extrapolated by \cite{shapiro14}
from the Sun to more active stars, we obtain \sindex~corresponding to a given light curve.

The ARs in our model consist of spot umbrae, spot penumbrae, and faculae. The contrast of a particular magnetic feature relative to the quiet regions (i.e. regions on the stellar surface free from any apparent manifestation of magnetic features) depends on the wavelength and the position of the magnetic feature on the stellar disc. We have used wavelength-dependent contrasts calculated by \cite{unruh99} and convolved them with the \emph{Kepler} spectral passband \citep[see][for a comprehensive study of the effect of different passbands on the 
measured variability]{nina20}.

The areas of the spot parts of ARs are randomly drawn from a log-normal probability 
distribution, which considers spot groups at their maximum-development 
phase, given by \citet[][Table 1]{bs05}. 
We choose areas to fall 
between 60 and 3000 $\mu$SH (millionths of a solar hemisphere), to cover the 
entire range of sunspot group areas in the part of the RGO record for which the distribution 
can be fit by a log-normal function. Following \cite{wenzleretl2006} we attribute 80\% of spot area to penumbra and the remaining 20\% to umbra. The ratio between facular and spot areas of ARs at the moment of emergence, $A_f/A_s$, is assumed to be the same for all ARs for a given stellar activity level \citep[see Fig.~10 from][]{shapiro20}. It is calculated to fulfil the relationship between the instantaneous facular and spot disc-area coverages observed for the Sun and extrapolated to more active stars \citep[namely, Eqs.~1--2 from][]{shapiro14}. 

%Lcurve-free-s19-decomp.pdf

\begin{figure*}
\centering
\includegraphics[width=.49\columnwidth]{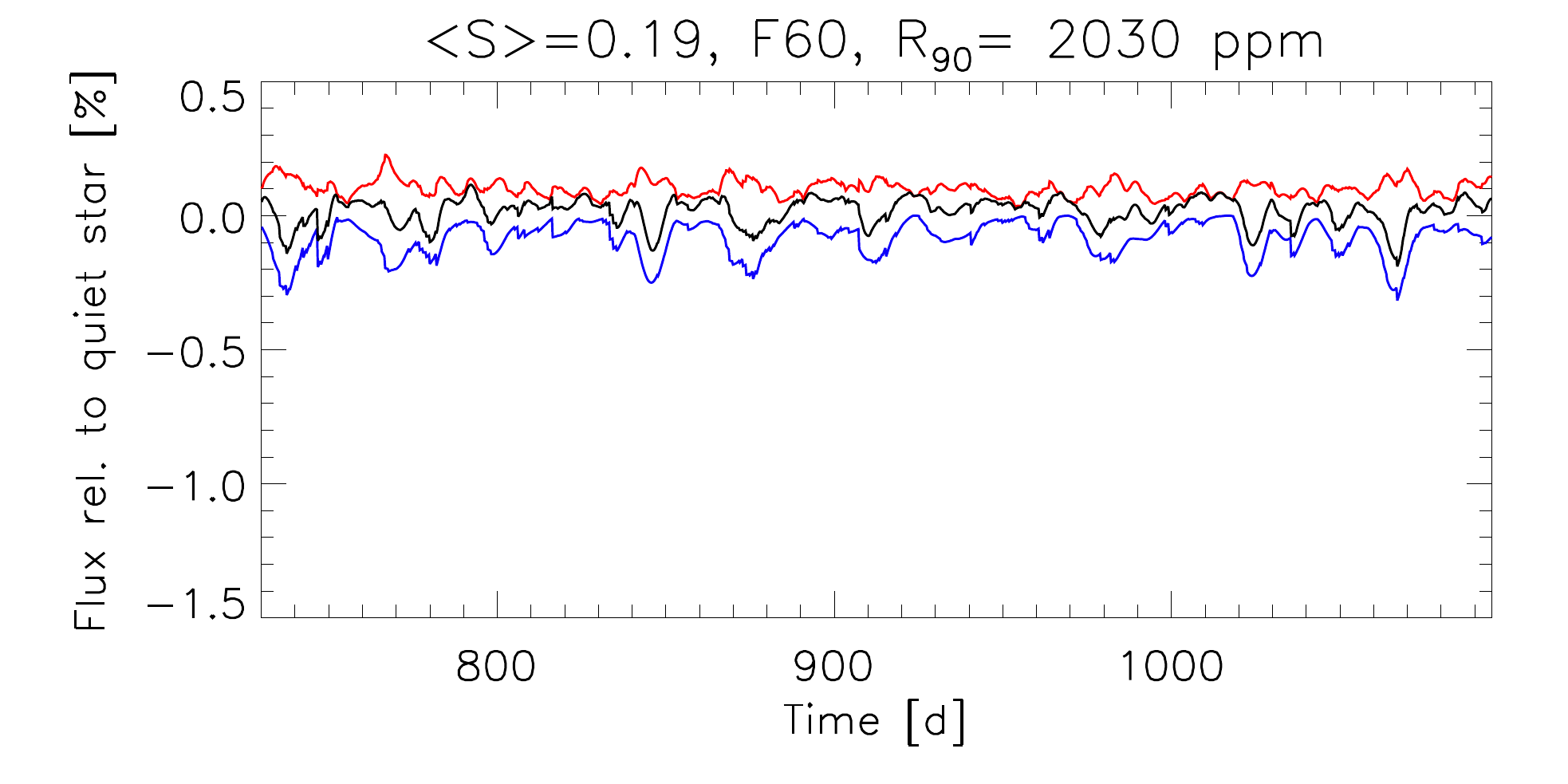}
\includegraphics[width=.49\columnwidth]{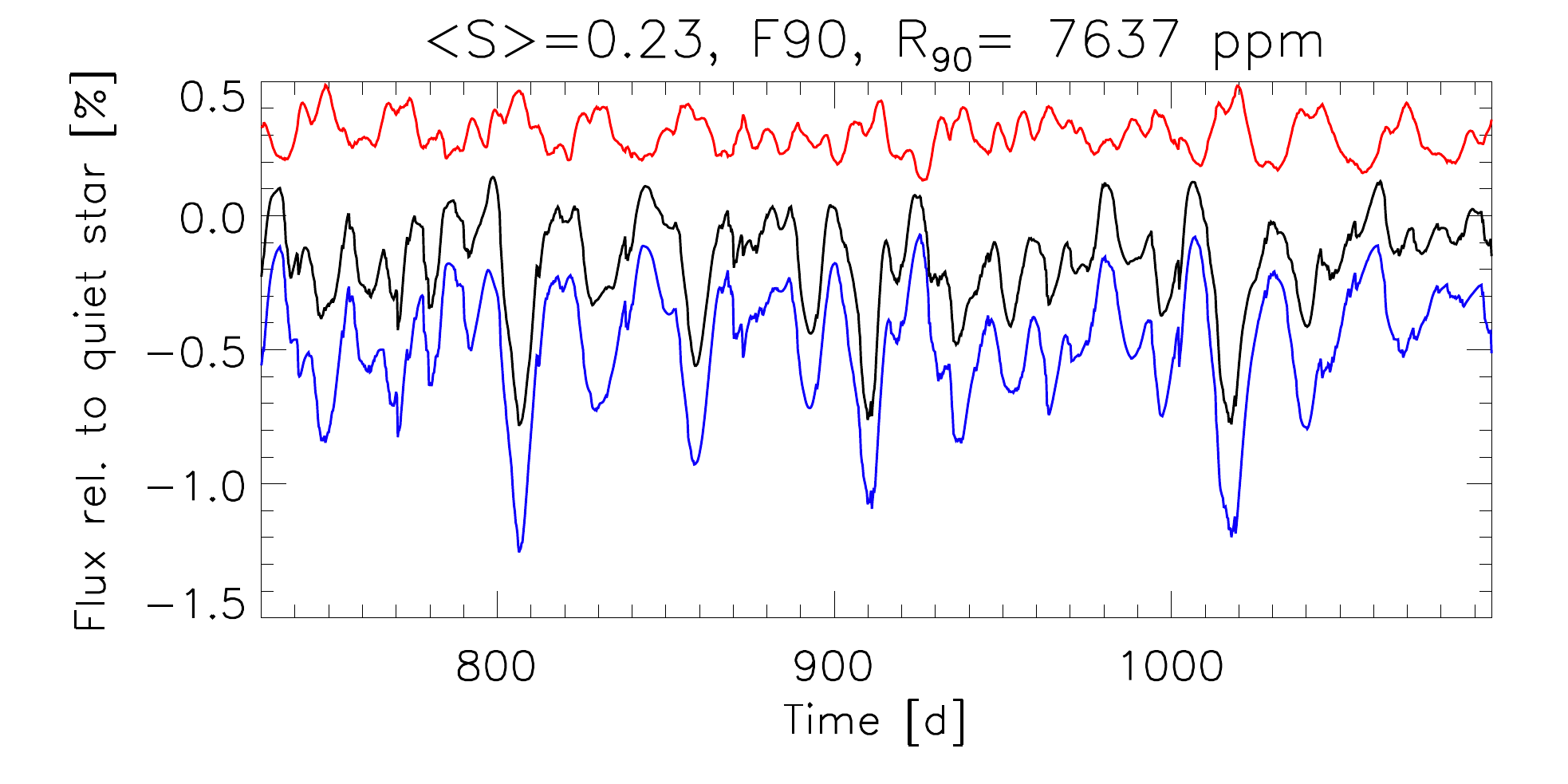} \\
\caption{{ Light curves (black) decomposed into facular (red) and spot (blue) components, for free-nesting mode. Left panel: low-activity, low-nesting ($S=0.19$, F60 case); right panel: high-activity, high-nesting ($S=0.23$, F90 case). The facular-to-spot area ratios during AR emergence are 2.92 and 1.43 for $S=0.19$ and $S=0.23$, respectively (see main text of the Appendix for details on how it is calculated).}}
\label{fig:Lcurved}
\end{figure*}

We do not consider a specific geometrical shape for an AR, prescribing the same value of the foreshortening to the entire region.  
This approximation is justified by the fact that the detailed 
morphology of ARs (i.e. spot group surrounded by facular regions) has a small effect on rotational variability, 
as long as the ARs are in the same size range with solar ARs, 
which we assume here for other solar-like stars. 

Following the emergence, we model the decay of 
spots and faculae as a continuous reduction of their areas. 
We consider a linear decay law following an instantaneous emergence, 
with a spot decay rate of $25~\mu$SH~d$^{-1}$, given that quadratic and linear decay laws are not 
distinguishable in the RGO dataset \citep{bs05}. For a given active region a decay time of its facular part was set to be three times longer than those of its spot part \citep[see][for a detailed discussion and case study]{shapiro20}.

The input physical parameters 
determining the final light curve are listed in Table~\ref{tab:pars}. 

{ Two example light curves are shown in Fig.~\ref{fig:Lcurved} (taken from Fig.~\ref{fig:LCfree}), in which the spot and facular components are shown separately. The facular-to-spot area ratio during the emergence of any spot is determined by the S-index, such that the ratio decreases with $S$ (see above). As a result, the relative contribution of the spot component in the total variability increases with the activity level. }
\begin{table}
\centering
\caption{Model parameters}
\label{tab:pars}
\begin{tabular}{ll}
\hline\hline
Parameter & Value \\
\hline
Equatorial rotation period & 25~d~(equatorial) \\
Axial inclination & $90^\circ$ \\
Spot area distribution & Log-normal\tablenotemark{a} \\
Spot area range & $60<A_s<3000~\mu$SH \\
%Area ratio (emergence) & $A_f/A_s=3$ \\
Spot decay law & Linear;~$25~\mu$SH~d$^{-1}$ \\
Facula-spot lifetime ratio & $\tau_f/\tau_s=3$ \\
%Activity level\tablefootmark{b} & 1350 & Variable \\
%Rotation profile & Variable & Rigid in latitude\tablenotemark{d} \\
%Nesting degree & Variable & 0.0 \\
\hline
\end{tabular}
%\tablenotetext{a}{See main text for definitions.}
%\tablenotetext{a}{Whether the parameter is fixed or allowed to vary.}
%\tablefoottext{b}{Number of spots (and faculae) to emerge during 4.4 years.}
\tablenotetext{a}{\citet{bogdan88,bs05}}
%\tablenotetext{d}{In case of differential rotation, the profile is from \citet{snodgrass83}}
\end{table}

\bibliography{clumpy}{}
\bibliographystyle{aasjournal}

%% Include this line if you are using the \added, \replaced, \deleted
%% commands to see a summary list of all changes at the end of the article.
%\listofchanges

\end{document}